# Colossal Magnetoresistance by Avoiding a Ferromagnetic State in the Mott System $Ca_3Ru_2O_7$

X.N. Lin[1], Z.X. Zhou[2], V. Durairaj[1], P. Schlottmann[2] and G. Cao[1*]

[1]Department of Physics and Astronomy, University of Kentucky, Lexington, KY 40506

[2] National High Magnetic Field Laboratory, Florida State University,

Tallahassee, FL 32310

Transport and magnetic studies of $Ca_3Ru_2O_7$ for temperatures ranging from 0.4 K to 56 K and magnetic fields, B, up to 45 T leads to strikingly different behavior when the field is applied along the different crystal axes. A ferromagnetic (FM) state with full spin polarization is achieved for B||a-axis, but colossal magnetoresistance is realized *only* for B||b-axis. For B||c-axis, Shubnikov-de Haas oscillations are observed and followed by a less resistive state than for B||a. Hence, in contrast to standard colossal magnetoresistive materials, the FM phase is the *least favorable* for electron hopping. These properties together with highly unusual spin-charge-lattice coupling near the Mott transition (48 K) are driven by the orbital degrees of freedom.

In ruthenates, in particular the bi-layered $Ca_3Ru_2O_7$, the orbital degrees of freedom play an important role prompting novel phenomena through the coupling of the orbits to the spin (spin-orbit interaction) and to the lattice (Jahn-Teller effect). The extension of the 4d-orbitals leads to comparable and thus competing energies for the crystalline fields, Hund's rule interactions, spin-orbit coupling, p-d hybridization and electron-lattice coupling. Physical properties are then particularly susceptible to small perturbations by external magnetic fields and slight structural changes. A colossal magnetoresistance (CMR) normally occurs for the field applied along the easy axis of magnetization. In this letter we present evidence for an unusual CMR in $Ca_3Ru_2O_7$ realized *only* when B is *perpendicular* to the easy axis of magnetization (*a*-axis) or when the spin-polarized state is avoided. This CMR phenomenon is fundamentally different from those of all other magnetoresistive systems, which are primarily driven by spin polarization. In addition, for B||*c*-axis, Shubnikov-de Haas oscillations are observed and followed by a much more conducting state than that for B||*a*. The results in this letter include the resistivity for the current along the c-axis, $\rho_c$, taken in B up to 45 T at 0.4 K and the field dependence of the magnetization, M, and $\rho_c$ in the vicinity of $T_{MI}$ (40≤T<56 K) for B applied along the *a*-, *b*-, and *c*-axis. All results provide a coherent picture illustrating that the spin-polarized state is energetically the *least favorable* for electron hoping, and that orbital order and its coupling to lattice and spin degrees of freedom drive the exotic electronic and magnetic properties in $Ca_3Ru_2O_7$.

In ruthenates with $Ru^{4+}$ ($4d^4$) ions, the Hund's rule energy maximizing the total spin at each Ru site is not large enough to overcome the $e_g$-$t_{2g}$ crystalline field splitting, so that the $e_g$ levels are not populated. Hence, one $t_{2g}$ orbital is doubly occupied, while the other two host a single electron each. The three $t_{2g}$ levels ($d_{xy}$, $d_{zx}$ and $d_{yz}$) are expected to have different energies because the RuO$_6$ octahedra are deformed (all lattice parameters are different). These splittings are larger than the thermal energy k$_B$T and the Zeeman effect. The octahedra are corner-shared and often tilted. The tilting plays an important role in the overlap between orbitals of neighboring octahedra. Small changes in the tilting can result in qualitative changes in the properties. Local density approximation calculations for Sr$_3$Ru$_2$O$_7$, which has common aspects with Ca$_3$Ru$_2$O$_7$, find the Fermi surface very sensitive to small structural changes which readily shift the Fermi energy [1]. In view of the strong sensitivity to crystalline field and tilting angle asymmetries, the coupling of the magnetic field to the system depends on the orientation of the field. Consequently, different properties can be expected if the magnetic field is applied along the different crystallographic axes.

The Mott transition in the single-layered Ca$_{2-x}$Sr$_x$RuO$_4$ has been attributed theoretically [2-5] and experimentally [6, 7] to Coulomb interactions and orbital ordering (OO). A wealth of experimental results for the bi-layered Ca$_3$Ru$_2$O$_7$ still needs interpretation but their origin is likely to be driven by OO [8-19]. The crystal structure of Ca$_3$Ru$_2$O$_7$ is severely distorted by a tilt of the RuO$_6$ octahedra [9], which projects primarily onto the *ac*-plane (153.22°), while it only slightly affects the *bc*-plane (172.0°) [9]. These crucial bond angles directly impact the band structure and are the origin of the anisotropic properties of the compound. In zero-field Ca$_3$Ru$_2$O$_7$ undergoes an antiferromagnetic (AFM) transition at T$_N$=56 K while remaining metallic, and then a Mott-like transition at T$_{MI}$=48 K [8-19] with a dramatic reduction (up to a factor of 20) in the conductivity for T < T$_{MI}$ [8-12]. This transition is accompanied by an abrupt shortening of the *c*-axis lattice parameter below $T_{MI}$ [10]. Such magnetoelastic coupling results in Jahn-Teller distortions of the RuO$_6$ octahedra [9, 10], thus lowering $d_{xy}$ orbitals relative to d$_{zx}$ and $d_{yx}$ orbitals with a possible orbital distribution of $(n_{xy}, n_{zx}/n_{yz})=(2,2)$ [18]. Consequently, an AFM and OO phase can occur, explaining the poor metallic behavior for $T<T_{MI}$ and $B<B_c$ (critical field for the metamagnetic transition). This is consistent with Raman-scattering studies of Ca$_3$Ru$_2$O$_7$ revealing that the transition at $T_{MI}$ is associated with the opening of a charge gap, $\Delta_c \sim 0.1$ eV, in one of the bands, and the concomitant softening and broadening of an out of phase O phonon mode. The result can be explained by a rearrangement in the $t_{2g}$ orbitals [13, 14, 18].

Single crystals were grown using both flux and floating zone techniques [20] and characterized by single crystal x-ray diffraction, Laue x-ray diffraction, SEM and TEM. All results indicate that the single crystals are of high quality. The highly anisotropic magnetic properties of Ca$_3$Ru$_2$O$_7$ are used to determine the magnetic easy *a*-axis and to identify twinned crystals that often show a small kink at 48 K in the *b*-axis susceptibility.

Shown in Fig.1 is the field dependence of the resistivity for the *c*-axis (inter-plane) $\rho_c$ (right scale) for T=0.4 K and 0≤B≤45 T with B||*a*-, *b*- and *c*-axis (the *a*-axis resistivity, $\rho_a$, exhibits a similar field dependence, and is not shown here). $\rho_c$ is extraordinarily sensitive to the orientation of B. For B||*a*-axis *(magnetic easy-axis)* $\rho_c$ shows an abrupt drop by an order of magnitude at 6 T corresponding to the first-order metamagnetic transition leading to the spin-polarized or FM state with a saturated moment, M$_s$, of 1.8 µ$_B$/Ru or more than 85% polarized spins (see left scale in Fig.1) [8-9]. The reduction of $\rho_c$ is attributed to the coherent motion of electrons between Ru-O planes separated by insulating Ca-O planes, a situation similar to spin-filters where the probability of tunneling depends on the angle between the spin magnetization of adjacent ferromagnets. The same reduction in ρ is also observed for the current along the *a*- and *b*-axis, suggesting that the magnetoresistivity is driven entirely by the in-plane spin-reorientation for B||*a*-axis.

The fully spin polarized state can lower the resistivity by at most a factor of 10. As B is increased further from 6 to 45 T, $\rho_c$ increases linearly with B by more than 30%. A linear variation of $\rho_c$ over such a wide range of B is interesting in its own right, since a quadratic dependence is expected for regular metals [21]. Because spin scattering is already reduced to its minimum at B=6 T, the linear increase can only arise from orbital degrees of freedom which via spin-orbit coupling hinder the electrons from hopping.

For B||*b*-axis (magnetic hard-axis) there is no spin-flop transition and the system remains AFM. In sharp contrast to $\rho_c$ for B||*a*-axis, $\rho_c$ for B||*b*-axis rapidly decreases by as much as three orders of magnitude at $B_c$=15 T, two orders of magnitude more than that for B||*a*, where spins are fully polarized. Here $B_c$ decreases with increasing T. For B||*c*-axis*,* on the other hand, $\rho_c$ displays Shubnikov-de Haas (SdH) oscillations with frequencies of 28 T and 10 T, corresponding to very small Fermi surface cross sections [22] (The SdH effect re-appears with vastly different behavior when B rotates within the *ac*-plane [23]). Remarkably, $\rho_c$ for $B_{||c}$>39 T is much smaller than $\rho_c$ for B||*a*. From 0 to 45 T, $\rho_c$ decreases by a factor of 7 and 40 for B||*a* and B||*c*, respectively. It is striking that a fully spin-polarized state, which is essential for magnetoresistance in all other magnetoresistive materials [24, 25], is the *least favorable* for conduction in $Ca_3Ru_2O_7$.

Shown in Fig. 2 are the magnetization M and $\rho_c$ as a function of B for B||*a*- and *b*-axis for 40≤T<56 K. The advantage of this temperature range is that $B_c$, along both the *a*-

and *b*-axis, falls into the range of 7 T, so that M can be fully characterized using a SQUID magnetometer. A direct comparison of ρ and M allows to further probe correlations and the role of OO. Fig. 2a displays M as a function of B for B||*a*-axis. At 40 K, M(B) is still very similar to M(B) at low temperatures (see Fig.1) but with slightly lower $M_s$ (=1.6 $\mu_B$/Ru) and $B_c$ (=5.8 T). For 41≤T≤45 K, a second transition develops at B*>$B_c$, suggesting an intermediate FM state for $B_c$<B<B*, which is not fully polarized along the *a*-axis. A possible interpretation is that the spins are rotating away from the *a*-axis due to a shortening of the c-axis near $T_{MI}$ and hence a stronger field (B*) is required to re-align these spins along the *a*-axis. Since the spin rotation tends to become stronger as T approaches $T_{MI}$, B* increases with T. M is about 1 $\mu_B$/Ru at $B_c$ and increases by 0.6 $\mu_B$/Ru at B*. Only half of the ordered spins are thus aligned with the *a*-axis in the spin reorientation (SR) region for $B_c$<B<B*. $B_c$ decreases with T and vanishes near $T_N$ (=56 K). Unlike M for B||*a*-axis, M for B||*b*-axis is unsaturated at B>$B_c$ and rounded at $B_c$ without hysteresis, suggesting a second-order transition (see Fig. 2b). Noticeably, M at 7 T always converges to ~1 $\mu_B$/Ru for all temperatures, signifying that the spin polarization is roughly 50% along the *b*-axis and independent of T. Clearly, $M_s$ for B||*b*-axis is always smaller than that for B||*a*-axis in spite of the spin reorientation that partially enhances M for B||*b*-axis.

The corresponding $\rho_c$ as a function of B for B||*a*- and *b*-axis is displayed in Figs. 2c and 2d, respectively. For B||*a*-axis, $\rho_c$ at 40 K shows an abrupt drop at $B_c$ similar to that at low temperatures with a magnetoresistance ratio Δρ/ρ(0)=58%, where Δρ=ρ(7T)-ρ(0). In the range 41≤T≤45 K, $\rho_c$ for B||*a* decreases initially at $B_c$ and then further at B* with a total Δρ/ρ(0) similar to that at 40 K. Clearly, for T≤45 K $\rho_c$ perfectly mirrors the

behavior of M for B||*a*-axis, suggesting a strong spin-charge coupling in this region. However, for T ≥ 46 K a valley develops in $\rho_c$; the beginning and end of this valley defines two fields, $B_{c1}$ ($B_{c1}=B_c$ for T<46K) and $B_{c2}$. The valley broadens with increasing T ($B_{c1}$ decreases with T, while $B_{c2}$ increases) and changes its shape for T ≥ 48.2 K, where the slope at $B_{c1}$ is now positive and $B_{c1}$ increases with T. *An important point is that the field dependence of $\rho_c$ for 46≤T≤52 K does not track the field dependence of M* (compare Figs. 2a and 2c). This lack of parallel behavior of M and $\rho_c$ is precisely *a* manifestation of the crucial role of the orbital degrees of freedom that dictate electron hopping for B||*a*.

In contrast, for large B||*b*-axis (Fig. 2d) $\rho_c$ always decreases to a value lower than the corresponding one for B||*a*-axis at B>$B_c$. This is the case despite that $M_s$ for B||*b*-axis is always smaller than $M_s$ for B||*a*-axis. For example, at 42 K and 7 T, $\Delta\rho/\rho(0)$ =50% with $M_s$=1.52 $\mu_B$/Ru for B||*a*, and $\Delta\rho/\rho(0)$ =80% with $M_s$=1.03 $\mu_B$/Ru for B||*b*. Note that the difference in both $M_s$ and $\Delta\rho/\rho(0)$ between B||*a*- and *b*-axis is ~35%. The temperature dependence of M(7T) (left scale) and $\Delta\rho/\rho(0)$ at 7 T (right scale) for B||*a*- and *b*-axis is summarized in Fig. 3a. The inverse correlation between M and $\Delta\rho/\rho(0)$ illustrated suggests that the spin-polarized state is indeed detrimental to the CMR. For T>$T_{MI}$, the metallic state is recovered for B<$B_{c1}$. But applying B along the *a*-axis leads to a rapid increase in $\rho_c$ with positive $\Delta\rho/\rho(0)$ reaching as high as 112% for B>$B_{c2}$, whereas applying B along the *b*-axis results in essentially no changes in $\rho_c$.

As known, for B<6 T, the Mott-like state is due to the OO facilitated by the c-axis shortening at $T_{MI}$ [12,18]. When $B_{||a}$>6 T the magnetic state becomes FM with the OO remaining and stabilized by the FM state. The orbital order is either a ferro-orbital (FO) or an antiferro-orbital (AFO) configuration. Hence the system is in either FM/FO or FM/AFO state. The former inhibits the hopping of the 4d electrons because of the Pauli's exclusion principle, while the latter permits intersite transitions but at the expense of a Coulomb energy. Therefore, despite of an order of magnitude drop in $\rho_c$ due to the spin-polarization when B>6 T, a fully metallic state can never be reached for B||*a*-axis. In fact, the linear increase in $\rho_c$ with increasing B for $B_{||a}$>6 T as shown in Fig.1 may manifest a strengthened OO via the enhanced FM state. Conversely, applying B along the *b*-axis steadily suppresses the AFM state [12], removing the orbital order through spin-orbit interaction when B>$B_c$. Such an orbitally disordered (OD) state drastically increases the electron mobility, therefore leading to the CMR. On the other hand, applying B along the *c*-axis has a noticeable impact on spin and orbital configurations when B>35 T where $\rho_c$ drops rapidly and becomes much smaller than $\rho_c$ for B||*a*-axis. This suggests that the electronic state for B||*a*-axis is the most resistive one.

The magnetic and transport behavior shown in Fig.2 is remarkably consistent with rapid changes of the Ru-O phonon frequency with B seen in Raman studies (Fig. 2b in Ref. 18), providing complementary evidence for the evolution of the field-induced magnetic and orbital phases. While applying B along the *b*-axis clearly favors the CMR, applying B||*a*-axis generates a rich phase diagram (see Figs. 3b and 3c). As shown in Fig. 3b, below 40 K, B drives the system from an AFM/OO to a FM/OO state, and for 40<T<48 K the system enters a region of spin reorientation (SR) characterized by $B_c$ and B*. For 46≤T<48.2 K, the valley seen only in $\rho_c$ signals an onset of an OD state at $B_{c1}$ and then a re-occurring OO state at $B_{c2}$ characterized by a sharp increase in $\rho_c$. For 48.2<T<56 K, the system changes from an AFM/OD to an AFM/OO phase when B>$B_{c1}$.

The evolution of the magnetic/orbital configuration is associated with the Jahn-Teller coupling, which appears to be most drastic in the vicinity of $T_{MI}$.

We have presented evidence that the orbital degree of freedom and its coupling to spin and lattice play a critical role in $Ca_3Ru_2O_7$. As a consequence, applying B along the *a-*, *b-,* and *c*-axis leads to novel and vastly different properties. Most notably, the CMR achieved by avoiding the FM state is fundamentally different from that of all other magnetoresistive materials.

**Acknowledgements:** The authors thank Dr. L. Balicas for his help with measurements in the 45 T hybrid magnet at NHMFL. G.C. is grateful to Dr. S. Lance Cooper for his important comments on the paper and Dr. Ganpathy Murthy for very helpful discussions. This work was supported by NSF (grant No. DMR-0240813). P.S. acknowledges the support by NSF (grant No. DMR01-05431) and DOE (grant No. DE-FG02-98ER45707).

* Corresponding author: cao@uky.edu

**References:**

1. D.J. Singh and I.I. Mazin, Phys. Rev. B **63**, 165101 (2001).

2. T. Hotta and E. Dagotto, Phys. Rev Lett. **88**, 017201 (2002).

3. V.I. Anisimov, *et al*., Eur. Phys. J. B **25**, 191 (2002).

4. A. Liebsch, Phys. Rev. Lett. **91**, 226401 (2003).

5. Z. Fang, et al, Phys. Rev. B **69**, 045116 (2004).

6. L.S. Lee, *et al*., Phys. Rev. Lett. 89, 257402 (2002)

7. T. Mizokawa *et al.*, Phys. Rev. Lett. **87**, 077202 (2001).

8. G. Cao, *et al*., Phys. Rev. Lett. **78**, 1751 (1997).

9. G. Cao, *et al.,* Phys. Rev. B **62**, 998 (2000).

10. G. Cao, *et al.,* Phys. Rev. B **67** 060406 (R) (2003).

11. G. Cao, *et al*., Phys. Rev. B **67** 184405 (2003).

12. G. Cao, *et al.,* Phys. Rev. B **69**, 014404 (2004)

13. H.L. Liu, *et al*., , Phys. Rev. B **60**, R6980, (1999).

14. C.S. Snow, *et al*., Phys. Rev. Lett. **89**, 226401 (2002).

15. A.V. Puchkov, *et al*., Phys. Rev. Lett. **81**, 2747 (1998).

16. R.P. Guertin, *et al*., Solid State Comm. **107**, 263 (1998).

17. S. McCall, *et al*, Phys. Rev. B **67,** 094427 (2003).

18. J.F. Karpus, et al., Phys. Rev. Lett. **93,** 167205 (2004)

19. E. Ohmichi, *et al*., Phys. Rev. B. **70**, 104414 (2004)

20. There is *no difference* in the magnetic and transport properties and Raman spectra of crystals grown using flux and floating zone methods. Our studies on oxygen-rich $Ca_3Ru_2O_{7+\delta}$ show that the resistivity for the basal plane shows a downturn below 30 K, indicating brief metallic behavior.

21. A.B. Pippard, *Magnetoresistance in Metals* (Cambridge University Press, Cambridge, 1989)

22. G. Cao, *et al.*, New Journal of Physics 6, 159 (2004).

23. X.N. Lin, *et al*, unpublished

24. For example, E.Y. Tsymbal and D.G. Pettifor, p. 113, *Solid State Physics* v. 56, ed. Henry Ehrenreich and Frans Spaepen (Academic Press, New York, 2001)

25. For example, Yoshinori Tokura, *Colossal Magnetoresistive Oxides* (Gordon and Beach Science Publishers, Australia, 2000).

**Captions:**

**Fig.1.** Isothermal magnetization M for B||*a-, b- and c*-axis at T=2 K (left scale). Magnetic field dependence of the resistivity for the *c*-axis $\rho_c$ for B||*a-, b-* and *c*-axis at T=0.4 K (right scale). Note that the magnetic easy-axis is along the *a*-axis with spin polarization of more than 85%.

**Fig.2.** M and $\rho_c$ as a function of B for B||*a*-axis (panels (a) and (b)) and *b*-axis (panels (c) and (d)) for 40≤T<56 K. The solid arrows (empty arrows) indicate B* or $B_{c1}$ ($B_{c2}$

**Fig.3.** (a) Temperature dependence of M (triangles) and $\Delta\rho/\rho(0)$ (solid circles, right scale) at 7 T for B||*a-* and *b*-axis. (b) and (c) phase diagrams plotted as B vs T summarizing various phases for B||*a-* and *b*-axis, respectively. Note that in (b) $B_{c1}(\rho)$ and $B_{c2}(\rho)$ indicate the curves are generated based on $\rho$, and $B_c(M)$ on M.

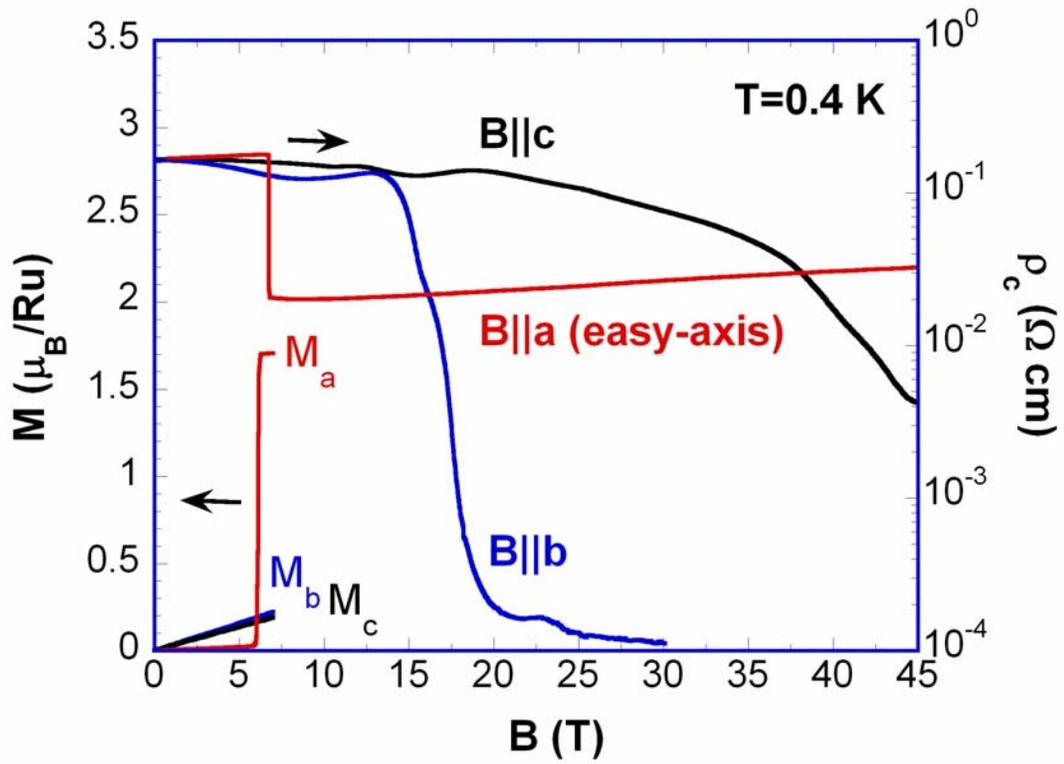

**Fig. 1**

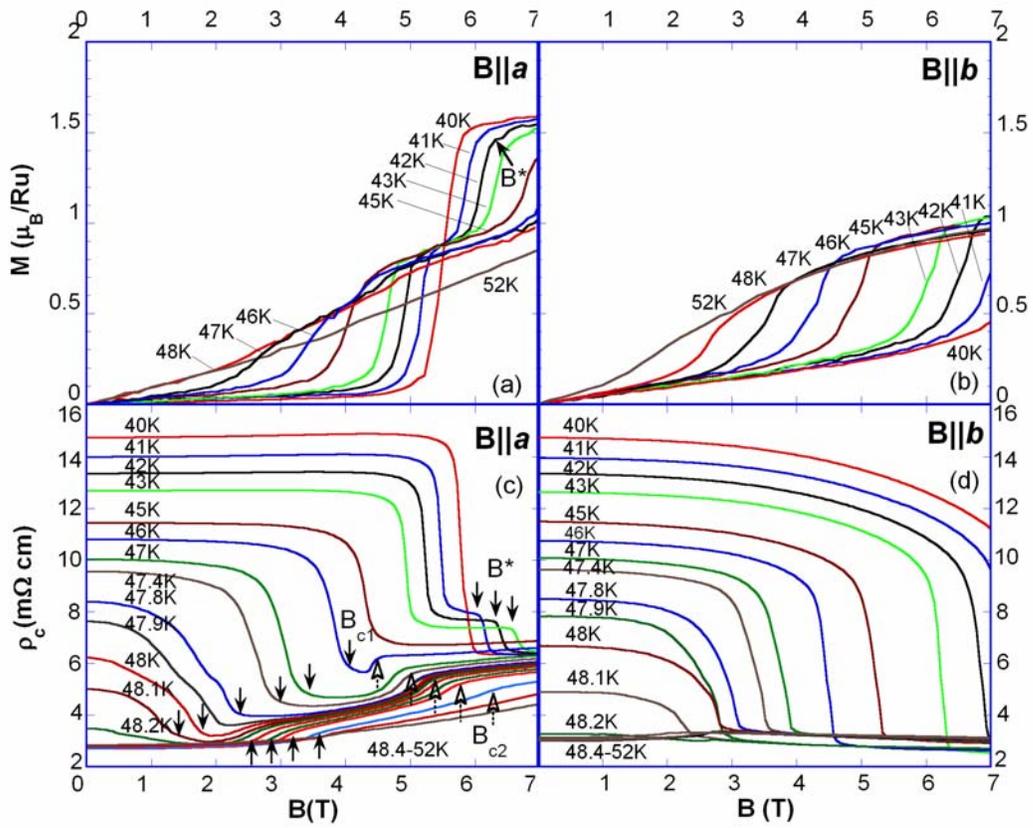

**Fig. 2**

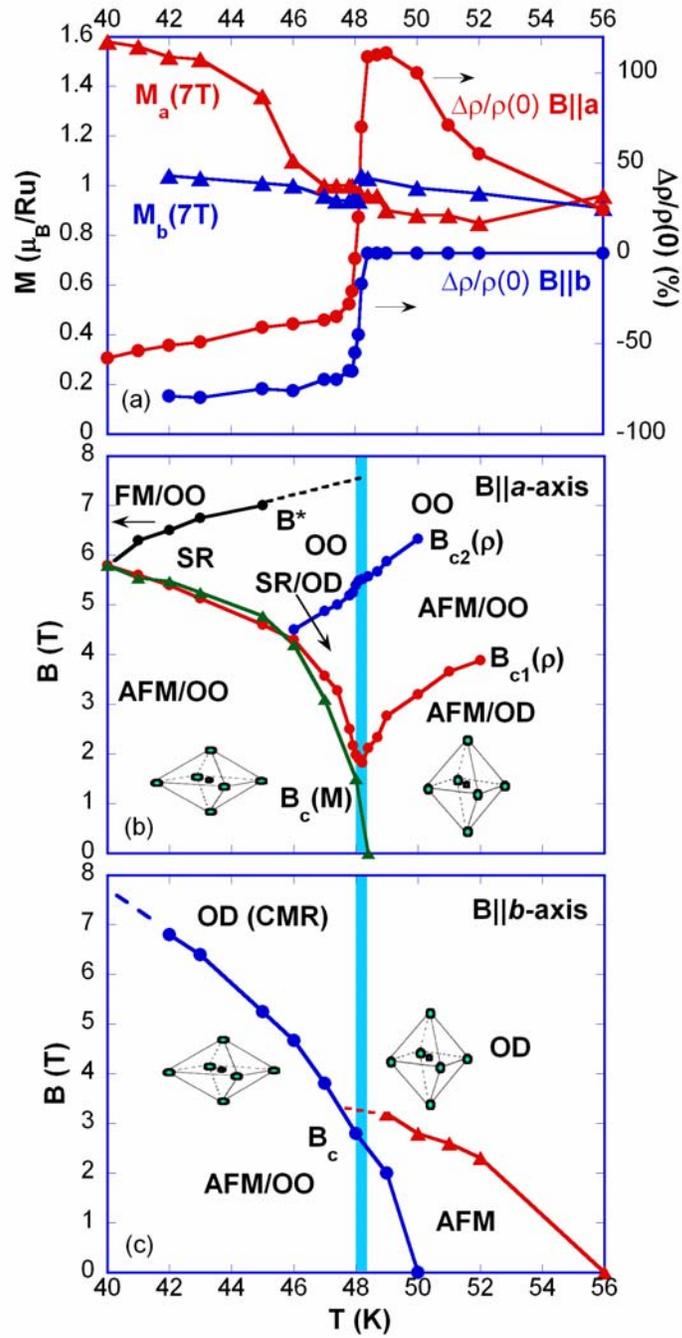

**Fig.3**